\documentclass[
showkeys,12pt,
preprint,preprintnumbers,nofootinbib,
groupedaddress,superscriptaddress,amsmath,amssymb]{revtex4}
\usepackage{graphicx}
\usepackage{dcolumn}
\usepackage{bm}
\usepackage{amssymb}
\usepackage{amsmath}
\usepackage{epsfig}    
\usepackage{color}
\usepackage{slashed}
\usepackage{hhline}

\def\be{\begin{equation}}
\def\ee{\end{equation}}
\newcommand{\bea}{\begin{eqnarray}}
\newcommand{\eea}{\end{eqnarray}}
\newcommand{\nn}{\nonumber}

\numberwithin{equation}{section}

\begin{document}

\title{
Two Loop Radiative Seesaw and X-ray line Dark Matter\\ with\\ Global $U(1)$ Symmetry
}
\preprint{KIAS-P15013}
\author{Hiroshi Okada}
\email{hokada@kias.re.kr}
\affiliation{School of Physics, KIAS, Seoul 130-722, Korea}
\affiliation{Physics Division, National Center for Theoretical Sciences, Hsinchu, Taiwan 300}

\date{\today}

\begin{abstract}
We study a two loop induced radiative neutrino model with global $U(1)$ symmetry at 1 GeV scale, in which we explain the X-ray line at 7.1 keV of  dark matter candidate reported by XMN-Newton X-ray observatory using data of various galaxy clusters and Andromeda galaxy.
We also discuss Higgs sector, lepton flavor violation processes, and a physical Goldstone boson.
\end{abstract}
\maketitle
\newpage

\section{Introduction}
One of the promising scenarios simultaneously to explain between Neutrinos and dark matter (DM) physics is to generate neutrino masses at multi-loop level~\cite{Ma:2006km,
Aoki:2013gzs, Dasgupta:2013cwa, Krauss:2002px, Aoki:2008av, Schmidt:2012yg, Bouchand:2012dx, Aoki:2011he, Farzan:2012sa, Bonnet:2012kz, Kumericki:2012bf, Kumericki:2012bh, Ma:2012if, Gil:2012ya, Okada:2012np, Hehn:2012kz, Dev:2012sg, Kajiyama:2012xg, Okada:2012sp, Aoki:2010ib, Kanemura:2011vm, Lindner:2011it, Kanemura:2011mw, Kanemura:2012rj, Gu:2007ug, Gu:2008zf, Gustafsson, Kajiyama:2013zla, Kajiyama:2013rla, Hernandez:2013dta, Hernandez:2013hea, McDonald:2013hsa, Okada:2013iba, Baek:2013fsa, Ma:2014cfa, Baek:2014awa, Ahriche:2014xra, Kanemura:2011jj, Kanemura:2013qva, Okada:2014nsa, Kanemura:2014rpa, Chen:2014ska, Ahriche:2014oda, Okada:2014vla, Ahriche:2014cda, Aoki:2014cja, 
Lindner:2014oea},~\cite{Ahn:2012cg, Ma:2012ez, Kajiyama:2013lja, Kajiyama:2013sza, Ma:2013mga, Ma:2014eka, Kanemura:2015mxa, Bahrami:2015mwa, Baek:2015mna, Hatanaka:2014tba, Okada:2014nea, Sierra:2014rxa, Okada:2014qsa, Okada:2014oda, Boehm:2006mi, Farzan:2014aca},~\cite{MarchRussell:2009aq}, in which DM could be a mediated field in the neutrino loop.
  
As for the DM sector, X-ray line signal at 3.55 keV from the analysis of XMN-Newton X-ray observatory data of various galaxy clusters and Andromeda galaxy~\cite{Bulbul:2014sua, Boyarsky:2014jta} can be understood by decaying scenario, in which the DM mass should be 7.1 keV and mixing angle between DM and the active neutrinos should be $\sin^22\theta\approx10^{-10}$.
Due to these simple implications, many works have been studied~\cite{Finkbeiner:2014sja, Jaeckel:2014qea,Lee:2014xua, Kong:2014gea,
 Frandsen:2014lfa, Baek:2014qwa, Cline:2014eaa, Modak:2014vva,
 Babu:2014pxa, Queiroz:2014yna, Demidov:2014hka, Ko:2014xda,
 Allahverdi:2014dqa,  Kolda:2014ppa, Cicoli:2014bfa, Dudas:2014ixa, Choi:2014tva, Okada:2014zea, Chen:2014vna, Conlon:2014xsa, Robinson:2014bma, Liew:2014gia, Chakraborty:2014tma, Tsuyuki:2014aia, Dutta:2014saa, Chiang:2014xra, Geng:2014zqa, Ishida:2014fra,  Baek:2014poa, Agrawal:2015tfa,
Lee:2015sha,
Arcadi:2014dca,
Kang:2014mea,
Patra:2014sua,
Harada:2014lma,
Cheung:2014tha,
Iakubovskyi:2014yxa,
Baek:2014sda,
Adulpravitchai:2014xna,
Patra:2014pga,
Faisel:2014gda,
Farzan:2014foo,
Jeltema:2014qfa,
Cline:2014kaa,
Haba:2014taa,
Biswas:2015sva,
Humbert:2015epa,
Merle:2014xpa,
Merle:2015oja,
Kusenko:2006rh,
Petraki:2007gq,
Merle:2013wta}.
\footnote{On the other hand, there are refuting papers~\cite{Carlson:2014lla, Jeltema:2014mla, Phillips:2015wla} that the 3.55 keV line may imply atomic transitions in helium-like potassium and chlorine are more likely to be the emitters.}

In our paper, we propose a two loop induced radiative neutrino model with a global $U(1)$ symmetry, in which such a small mixing between DM and neutrinos can be generated at one-loop level. Since the DM mass is expected to be very tiny, it does not link to the neutrino masses.   
Observed relic density can be thermally obtained by the annihilation process with a Goldstone boson (GB) pair that is a consequence of the global $U(1)$ symmetry.


This paper is organized as follows.
In Sec.~II, we show our model building including Higgs sector, neutrino masses, and lepton flavor violations (LFVs).
In Sec.~III, we analyze the DM properties including relic density and X-ray line.
In Sec.~IV, we numerically analyze the allowed region to satisfy all the conditions.
We conclude and discuss in Sec.~V.


\section{ Model setup}
 \begin{widetext}
\begin{center} 
\begin{table}[tbc]
\begin{tabular}{|c||c|c|c||c|c|c|c|}\hline\hline  
&\multicolumn{3}{c||}{Lepton Fields} & \multicolumn{4}{c|}{Scalar Fields} \\\hline
& ~$L_L$~ & ~$e_R^{}$~ & ~$N_R$ ~ & ~$\Phi$~ & ~$\chi^+_1$~  & ~$\chi^{+}_2$~ & ~$\varphi$ \\\hline 
$SU(2)_L$ & $\bm{2}$  & $\bm{1}$ & $\bm{1}$ & $\bm{2}$ & $\bm{1}$ & $\bm{1}$ & $\bm{1}$ \\\hline 
$U(1)_Y$ & $-1/2$ & $-1$  & $0$ & $+1/2$& $+1$& $+1$ & $0$  \\\hline
$U(1)$ & $-x$ & $-x$   & $x/3$ & $0$& $2x$& $2x/3$  & $-2x/3$  \\\hline
\end{tabular}
\caption{Contents of lepton and scalar fields
and their charge assignment under $SU(2)_L\times U(1)_Y\times U(1)$.}
\label{tab:1}
\end{table}
\end{center}
\end{widetext}

We discuss a two-loop induced radiative neutrino model. 
The particle contents and their charges are shown in Tab.~\ref{tab:1}. 
We add {three} gauge singlet Majorana fermions $N_R$,
two singly-charged  singlet scalars ($\chi^+_1, \chi_2^{+}$), and a neutral singlet scalar $\varphi$ to the SM.
We assume that  only the SM-like Higgs $\Phi$ and $\varphi$ have vacuum
expectation values (VEVs), which are symbolized by $v/\sqrt2$ and $v'/\sqrt2$ respectively. 
We also introduce a global $U(1)$ symmetry, under which
$x\neq0$ is an arbitrary number of the charge under the $U(1)$ symmetry, and 
the neutrino masses are induced at the two loop level after the spontaneous $U(1)$ symmetry breaking by $2x$  as shown in Figure~\ref{fig:neut1} (as well as  Figure~\ref{fig:neut2}).
Notice here that SM-like Higgs $\Phi$ should be neutral under the $U(1)$ symmetry not to couple quarks to Goldstone boson through chiral anomaly to be consistent with the axion particle search. Otherwise its breaking scale should be very large( that is greater than $10^{12}$ GeV)
, which is inconsistent with the observed value $v\approx246$ GeV.

The renormalizable relevant Lagrangian and Higgs potential under these symmetries are given by
\begin{align}
-\mathcal{L}_{Y}
&=
(y_\ell)_{ij} (\bar L_L)_i \Phi (e_R)_j  + f_{ij} \bar L_{L_i}^c\cdot L_{L_j} \chi_1^+  
+g_{ij} \bar N_{R_i} e_{R_j}^c \chi_2^- 
+ \frac12 (y_{N})_{ij} \varphi  \bar (N_R^c)_i (N_R)_j  +{\rm h.c.}, \\
{\cal V}&=
m^2_{\chi_1}|\chi_1^+|^2 + m^2_{\chi_2}|\chi_2^+|^2 + m^2_{\varphi}|\varphi|^2 + m^2_{\Phi}|\Phi|^2
+\lambda_0(\varphi^2 \chi_1^+ \chi_2^- +{\rm h.c.}) + \lambda_{\chi_1}|\chi_1^+|^4 + \lambda_{\chi_2}|\chi_2^+|^4 
\nn\\
&
+ \lambda_{\varphi}|\varphi|^4 + \lambda_{\Phi}|\Phi|^4
+\lambda_{\chi_1\chi_2}|\chi_1^+|^2|\chi_2^+|^2  +\lambda_{\chi_1\varphi}|\chi_1^+|^2|\varphi|^2 +\lambda_{\chi_1\Phi}|\chi_1^+|^2|\Phi|^2  \nn\\
&
+\lambda_{\chi_2\varphi}|\chi_2^+|^2|\varphi|^2  +\lambda_{\chi_2\Phi}|\chi_2^+|^2|\Phi|^2
 +\lambda_{\varphi\Phi}|\varphi|^2|\Phi|^2 ,
\label{Eq:lag-flavor}
\end{align}
where $i=1-3$, $j=1-3$, the first term of $\mathcal{L}_{Y}$ can generates the SM
charged-lepton masses $m_\ell\equiv y_\ell v/\sqrt2$ after the electroweak spontaneous breaking of $\Phi$, and we assume $\lambda_0$ to be real.
The Majorana mass ($M_N\equiv y_Nv'/\sqrt2$) is generated after the spontaneous breaking of $\varphi$.
The scalar fields can be parameterized as 
\begin{align}
&\Phi =\left[
\begin{array}{c}
w^+\\
\frac{v+\phi+iz}{\sqrt2}
\end{array}\right],\quad 
\varphi=\frac{v'+\sigma}{\sqrt{2}}e^{iG/v'},
\label{component}
\end{align}
where $v~\simeq 246$ GeV is VEV of the Higgs doublet, and $w^\pm$
and $z$ are respectively (non-physical) GB 
which are absorbed by the longitudinal component of $W$ and $Z$ boson.
Since the CP even bosons ($\phi,\sigma$) and singly-charged bosons $(\chi_1^+,\chi_2^+)$ mix each other through the term $|\Phi|^2|\varphi|^2$ and $\varphi^2 \chi_1^+ \chi_2^- $ respectively,
each of the resulting mass eigenstate and mixing matrix is reparameterized as~\cite{Okada:2014qsa} 
 \begin{align}
\left[\begin{array}{c} \sigma \\ \phi \end{array}\right]\equiv V^\dag\left[\begin{array}{c} h_1 \\ h_2 \end{array}\right],
\quad
\left[\begin{array}{c} \chi_1^+ \\ \chi_2^+ \end{array}\right]\equiv O^\dag \left[\begin{array}{c} h_1^+ \\ h_2^+ \end{array}\right],
\end{align}
where $h_2$ is the SM-like Higgs, $h_1$ is an additional CP-even Higgs mass eigenstate, $(h_1^+,h_2^+)$ is the singly charged boson mass eigenstate, and each of $O$ and $V$ is $2\times 2$ unitary mixing matrix. The lower bound of the singly-charged boson can be obtained by the LEP experiment; {\it i.e.}, 80 GeV $\lesssim m_{h^+_{1(2)}}$~\cite{pdf}.

\subsection{ Invisible decay of the SM-like Higgs ($h_2$)}
The current experiment at LHC suggests that the invisible branching ratio of the SM Higgs ($B_{\rm inv}$) is conservatively estimated to be less than 20 \% \cite{Belanger:2013kya, Garcia-Cely:2013nin}. There are three invisible modes: $h_2\to 2G$, $h_2\to 2N_k$, and $h_2\to 2h_1$, and their decay rates ($\Gamma_{\rm inv}$) are given by
\begin{align}
&\Gamma_{\rm inv}\equiv \Gamma(h_2\to 2G)+\Gamma(h_2\to 2N_k)+\Gamma(h_2\to 2h_1),\\
&\Gamma(h_2\to 2G) = \frac{m^3_{h_2}|V_{12}|^2}{32\pi v'^2},\\
&\Gamma(h_2\to 2N_k) = \frac{m_{h_2}M_{N_k}^2|V_{12}|^2}{16\pi v'^2} \left(1-\frac{4 M^2_{N_k}}{m^2_{h_2}}\right)^{3/2},\\
&\Gamma(h_2\to 2h_1) = \frac{|\mu_{h_2h_1h_1}|^2}{32\pi m_{h_2}} \sqrt{1-\frac{4 m^2_{h_1}}{m^2_{h_2}}},\\
& \mu_{h_2h_1h_1} = V^2_{11} (\lambda_{\varphi\Phi}V_{22} v +6 \lambda_{\varphi} V_{21} v')
+2 \lambda_{\varphi\Phi} V_{11}V_{12} (V_{21} v +V_{22} v')
+V^2_{12} (\lambda_{\varphi\Phi}V_{21} v' +6 \lambda_{\Phi} V_{22} v).
\end{align}
Then we have to satisfy the following relation
\begin{align}
\Gamma_{\rm inv}<\frac{B_{\rm inv} |V_{22}|^2}{1-B_{\rm inv}} \Gamma_{\rm Higgs}^{\rm SM},
\label{eq:Binv}
\end{align}
where $\Gamma_{\rm Higgs}^{\rm SM}\approx4$ MeV is the total decay width of the Higgs boson at $m_{h_2}=125$ GeV.

\subsection{ Constraint on the charged bosons}
\begin{figure}[tbc]
\begin{center}
\includegraphics[clip, scale=0.5]{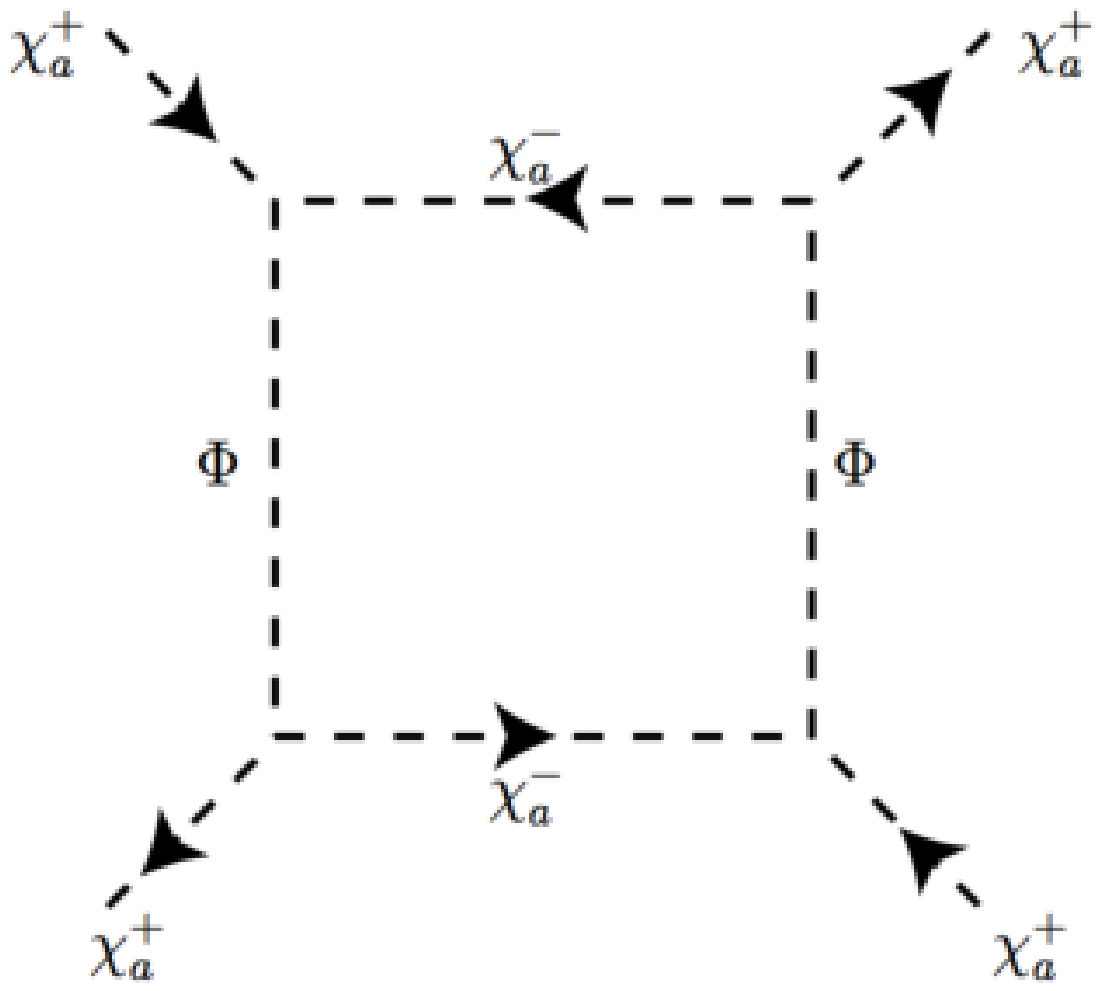}
\includegraphics[clip, scale=0.5]{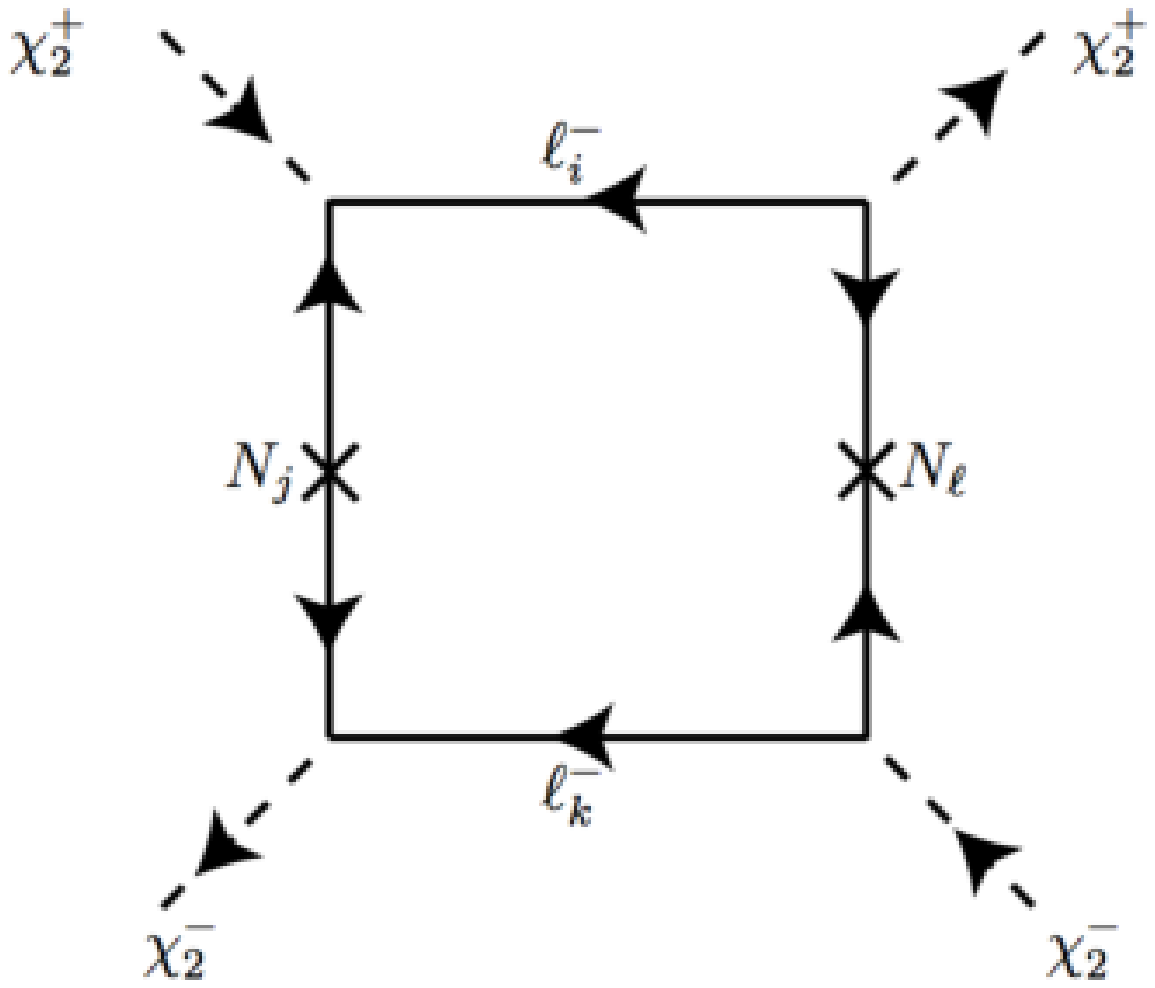}
\caption{One-loop contributions  for the pure quartic couplings of charged bosons $\lambda_{\chi_{1(2)}}$, where $a=1,2$ in the left-hand side.
}   \label{fig:4p-loop}
\end{center}
\end{figure}
The vacuum stability should be satisfied for the pure quartic couplings of charged bosons; $\lambda_{\chi_{1(2)}}$.
In our model, there exist some loop contributions to these couplings, by mediating bosons and fermions.
Here we consider this issue at the one-loop level.
As a general statement, we obtain a negative contribution relative to the tree-level coupling in case of boson-loop.
On the other hand, a relatively positive contribution is given  in case of fermion-loop.
We have only a boson-loop contribution to the $\lambda_{\chi_1}$ coupling as shown in the left-hand side of  Figure~\ref{fig:4p-loop},
on the other hand, we have both contributions to the $\lambda_{\chi_2}$ coupling as shown in  the right-hand side of Figure~\ref{fig:4p-loop}.
Then each of the condition up to the one-loop level  can be given as
\begin{align}
\lambda_{\chi_a}^{\rm one-loop}\approx
\lambda_{\chi_a}
&-\frac{3}{32(4\pi)^2}
\frac{(\lambda_{\chi_{\chi_a}\Phi}v)^4}{(M^2_{\chi_a}-M^2_{\Phi})^3} 
\left[
(M^2_{\chi_a} + M^2_{\Phi})
\ln\left[\frac{M^2_{\chi_a}}{M^2_{\Phi}}\right] - 2(M^2_{\chi_a}-M^2_{\Phi})
\right]\nn\\
&+
\delta_{a,\chi_2}\sum_{i,j,k,l}\frac{4M_{N_j}M_{N_l} }{(4\pi)^2}(g_{kj}^{\dag} g_{lk} g_{li} g_{ij}^{\dag})
\int\frac{dadbdcdd\delta(a+b+c+d-1)}{am_{\ell_k}^2+b m_{\ell_i}^2+c M_{N_\ell}^2+d M_{N_j}^2}\nn\\
&\gtrsim0,
\label{eq:vs}\\
M_{\chi_a}&\equiv 
m_{\chi_a} + \frac12(\lambda_{\chi_a\varphi} v'^2 + \lambda_{\chi_a\Phi} v^2),\quad
M_{\Phi}\equiv m_{\Phi} + \frac12 \lambda_{\varphi\Phi} v'^2=-\lambda_\Phi v^2,
\end{align}
where $a=1,2$, we have used the tadpole condition of $\Phi$ for the last term, and we neglect terms that are proportional to $v'^4$ or $(vv')^2$ because $v'(\approx{\cal O}(1)\ {\rm GeV})$ is very smaller than the mass scale of $M_{\chi_a(\Phi)}$ (that is ${\cal O}(100)$ GeV).\footnote{
The perturbativity and avoiding the global minimum can be straightforwardly satisfied  if  each of quartic coupling does not exceeds $4\pi$ and 
each of the sum of mass terms and the couplings are positive.}
The second term of the right-hand side in Eq.(\ref{eq:vs}) corresponds to the contribution of the boson-loop diagram as shown in  the left-hand side of  Figure~\ref{fig:4p-loop}, and The third term of the right-hand side in Eq.(\ref{eq:vs}) corresponds to the contribution of the fermion-loop diagram as shown in  the right-hand side of  Figure~\ref{fig:4p-loop}.


\subsection{ Neutrino mass matrix}
\begin{figure}[tbc]
\begin{center}
\includegraphics[clip, scale=0.5,angle=-90]{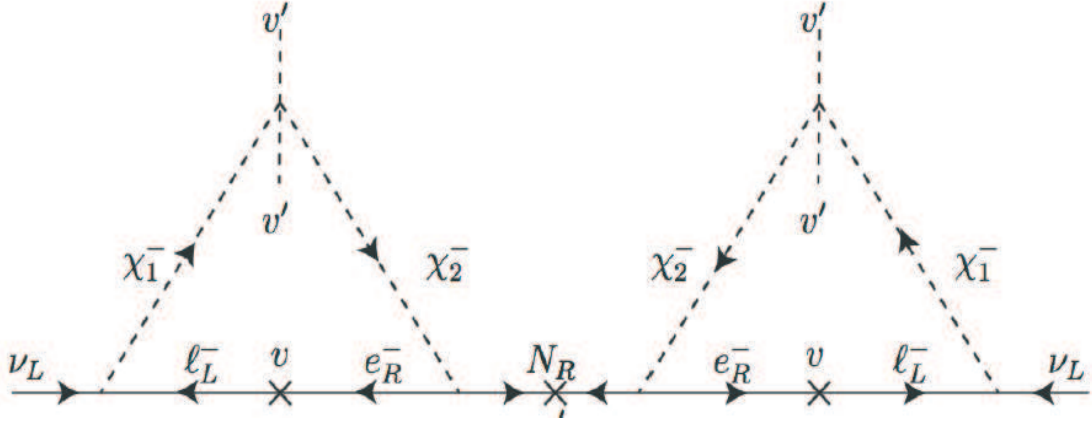}
\caption{Dominant contribution to the neutrino masses at the two-loop level.
}   \label{fig:neut1}
\end{center}
\end{figure}
At first we redefine relevant terms in terms of the mass eigenstate as
\begin{align}
\mathcal{L}_{Y}
&\sim
 f_{ij} (O^\dag)_{1a} \bar L_{L_i}^c\cdot L_{L_j} h_a^+  
+g_{ij}  (O^\dag)_{2a} \bar N_{R_i} e_{R_j}^c h_a^-  + \frac{y_N V_{1b}}{2\sqrt2} h_b^0 \bar N_R^c N_R   +{\rm h.c.}\nn\\
&\equiv
 f_{ij}^a \bar L_{L_i}^c\cdot L_{L_j} h_a^+  
+g_{ij}^a \bar N_{R_i} e_{R_j}^c h_a^-  + \frac{y_N^b}{2} h_b^0 \bar N_R^c N_R   +{\rm h.c.},
\label{Eq:lag-mass}
\end{align}
where $ f_{ij}^a\equiv  f_{ij} (O^\dag)_{1a}$, $ g_{ij}^a\equiv  g_{ij} (O^\dag)_{2a}$, and $ y_N V_{1b}/\sqrt2\equiv  y_N^b$.

Then the dominant contribution to the active neutrino mass matrix $m_\nu$  is given at two-loop level as shown in Figure~\ref{fig:neut1}, and its formula is given by 
\begin{align}
&(m_{\nu})_{ij}
=-(m_D)_{ik} M_{N_k}^{-1} (m_D^T)_{kj},\\
(m_D)_{ik}
=&
\frac{1}{(4\pi)^2}\sum_{a}^{1,2}\sum_{j}^{1-3}
f^a_{ij}m_{\ell_j}  g_{kj}^a \frac{\ln\epsilon_{aj}}{1-\epsilon_{aj}},
\end{align}
where $\epsilon_{aj}\equiv (m_{\ell_j}/m_{h_a^+})^2$, $M_{N_k}\equiv(M_{N_1},M_{N_2},M_{N_3})$, and $m_\ell=(m_e,m_\mu,m_\tau)$.
Notice here that another contribution to the neutrino mass matrix (Zee-Babu-like diagram as shown in Figure~\ref{fig:neut2}) can be tiny enough to be neglected in our case.
This formula is found in the {\it Appendix}.

The observed mixing matrix; PMNS(Pontecorvo-Maki-Nakagawa-Sakata) matrix ($U_{\rm PMNS}$)~\cite{Maki:1962mu}, can always be realized by introducing the Casas-Ibarra parametrization~\cite{Casas:2001sr},
where it is given by
\begin{widetext}
\begin{align}
\sum^{1-3}_{j}\sum^{1,2}_{a}f^a_{ij}F_{aj} (g^{aT})_{jk}
&=
U_{\rm PMNS}^* 
\left(\begin{array}{ccc}
m_{\nu_1}^{1/2} & 0 &0 \\ 0 & m_{\nu_2}^{1/2} &0 \\ 0 & 0 & m_{\nu_3}^{1/2} \\ 
\end{array}\right)
{\cal O} 
\left(\begin{array}{ccc}
M_{N_1}^{1/2} & 0 &0 \\ 0 & M_{N_2}^{1/2} &0 \\ 0 & 0 & m_{M_3}^{1/2} \\ 
\end{array}\right)
\left(\begin{array}{ccc}
m_{e}^{-1} & 0 &0 \\ 0 & m_{\mu}^{-1}  &0 \\ 0 & 0 & m_{\tau}^{-1}  \\ 
\end{array}\right)
,
\end{align}
\end{widetext}
$F_{aj}\equiv
\frac{1}{(4\pi)^2} \frac{\ln\epsilon_{aj}}{1-\epsilon_{aj}}$, and ${\cal O}$ is an arbitrary orthogonal matrix with complex values.
 Then the neutrino mass eigenvalues $m_\nu^{{\rm diag.}}\equiv(m_{\nu_1},m_{\nu_2},m_{\nu_3})$ can be given by
 \begin{align}
m^\dag_\nu m_\nu
&=U_{\rm PMNS}
\left(\begin{array}{ccc}
m_{\nu_1}^{2} & 0 &0 \\ 0 & m_{\nu_2}^{2} &0 \\ 0 & 0 & m_{\nu_3}^{2} \\ 
\end{array}\right)
U_{\rm PMNS}^\dag.
\end{align}
When $m_{h^+_a}= {\cal O}(100\ {\rm GeV})$, $F\approx0.01$ is obtained. Then $\sum_a f^a g^{aT}$ is written in terms of masses of neutrinos, charged-leptons, and right-handed neutrinos as
\begin{align}
\sum_a f^a g^{aT}\approx 10^2 \sqrt{\frac{m_\nu^{{\rm diag.}}M_N}{m^2_\ell}},
\end{align}
Depending on the charged-lepton masses, we can estimate a typical order of the Yukawa coupling as
\begin{align}
\sum_a f^a g^{aT} ={\cal O}(10^{-7}-10^{-3}),
\label{eq:yukawa-order}
\end{align}
where we fix $m_\nu^{{\rm diag.}}=0.01$ eV, and $M_N={\cal O}(10^{-5}-10^{-4})$ GeV.


\subsection{ Lepton Flavor Violations (LFVs) and lepton universalities}
\label{lfv-lu}
\subsubsection{$\ell_i^-\to\ell_j^+\ell_k^-\ell_l^-$ process}
The constraints from the $\ell_i^-\to\ell_j^+\ell_k^-\ell_l^-$ process~\cite{Beringer:1900zz} can be given at one-loop level as
 \begin{align}
& 
 \frac{
   \sqrt{
\left|(g_{b}^\dag g_{a})_{j\ell} (g_{a}^\dag g_{b})_{ik} + (g_{a}^\dag g_{b})_{jk} (g_{b}^\dag g_{a})_{i\ell}\right|^2
 +
 256
 \left|(f_{a}^\dag f_{a})_{j\ell} (f_{b}^\dag f_{b})_{ik} + (f_{a}^\dag f_{a})_{jk} (f_{b}^\dag f_{b})_{i\ell}\right|^2 
   }
 } 
 {8(4\pi)^2(m_{h_a^\pm}^2 -m_{h_b^\pm}^2)} 
 \ln\left[\frac{m_{h_a^\pm}^2}{m_{h_b^\pm}^2}\right]\nn\\
 &\hspace{5cm}
 \lesssim \frac{C_{ijk\ell}}{[{\rm GeV}]^2},
\end{align}
where each of $C_{ijk\ell}$ is given by
$C_{\mu eee}\approx2.3\times10^{-5}$, $C_{\tau eee}\approx0.009$, $C_{\tau ee\mu}\approx0.005$, $C_{\tau e\mu\mu}\approx0.007$, $C_{\tau \mu ee}\approx0.007$, $C_{\tau \mu e\mu}\approx0.007$, and $C_{\tau \mu\mu\mu}\approx0.008$~\cite{Herrero-Garcia:2014hfa}. Notice hereafter that we assume to be $m_\nu^2, M_{N}^2<<m_{h^\pm_{1(2)}}^2$. Notice here that the form $\frac{ \ln\left[{m_{h_a^\pm}^2}/{m_{h_b^\pm}^2}\right]} {m_{h_a^\pm}^2-m_{h_b^\pm}^2}$ reduces to $\frac{1}{m_{h_{b(a)}^\pm}^2}$ in the limit $a(b)\to b(a)$.


\subsubsection{$\ell_i^-\to\ell_j^-\gamma$ process}
The constraints from the $\ell_i^-\to\ell_j^- \gamma$ process~\cite{Adam:2013mnn} can be given at one-loop level as
 \begin{align}
\left| \left(\frac{(g_{a}^\dag g_{a})_{ij}}{4 m^2_{h^\pm_a}} \right)^2 + \left(\frac{(f_{b} f_{b}^\dag)_{ij}}{m^2_{h^\pm_b}} \right)^2
\right|
 \lesssim \frac{C_{ij}}{[{\rm GeV}]^4},
\end{align}
where each of $C_{ij}$ is given by
$C_{\mu e} \approx1.6\times10^{-6}$, $C_{\tau e}\approx0.52$, $C_{\tau \mu}\approx0.7$~\cite{Herrero-Garcia:2014hfa}.

\subsubsection{$\ell_i/\ell_j$ universality}
The constraint of the $\ell_i/\ell_j$ universality~\cite{Pich:2013lsa} is given as
\begin{align}
\left| \left| \frac{f_{ik}^a}{m_{h^\pm_a}}\right|^2 - \left| \frac{f_{jk}^a}{m_{h^\pm_a}}\right|^2 \right|
<\frac{C_{i/j}}{[{\rm GeV}]^2},
\end{align}
where $i\neq j\neq k$, and each of $C_{i/j}$ is given by $C_{\mu/e}\approx0.024$, $C_{\tau/\mu}\approx0.035$, and $C_{\tau/e}\approx0.04$~\cite{Herrero-Garcia:2014hfa}. 


\section{ Dark Matter}
\begin{figure}[tbc]
\begin{center}
\includegraphics[clip, scale=0.3]{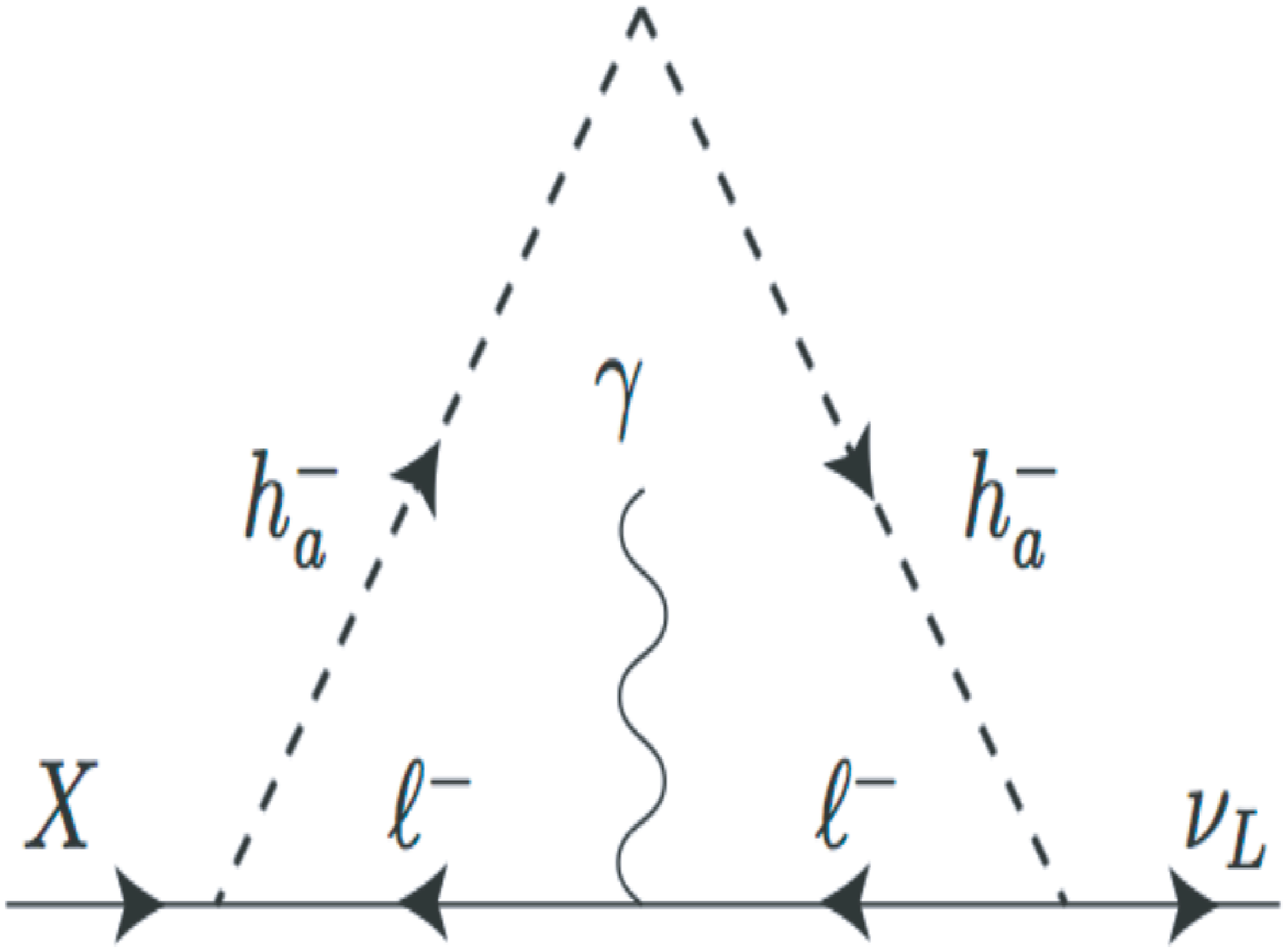}
\includegraphics[clip, scale=0.3]{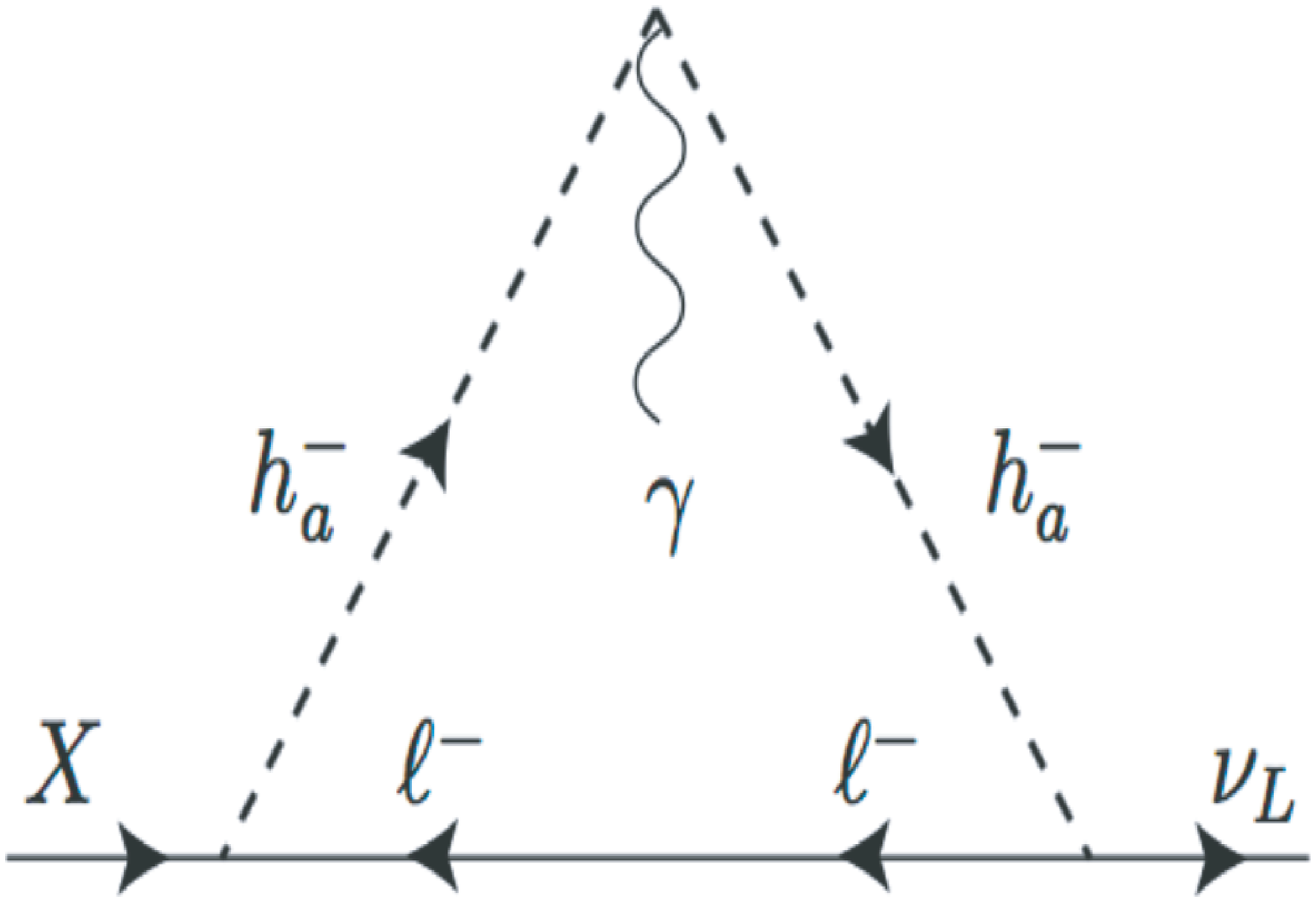}
\caption{The dominant contributions of the $X\to \nu_L\gamma$ process to explain the X-ray line. 
}   \label{fig:x-ray}
\end{center}
\end{figure}
We consider a fermionic DM candidate $X(\equiv N_1)$, which is assumed to be the lightest particle of $N_{i}$.
And we focus on the explanation of the X-ray line at 3.55 keV, since $X$ decays into active neutrinos and photon at the one-loop level after the spontaneous $U(1)$ symmetry breaking as shown in Fgiure~\ref{fig:x-ray}.
Then the mass of DM $M_X(\equiv M_{N_1})$ is fixed to be around 7.1 keV with a small value of the decay rate divided by $M_X$;
{\it i.e.}, 
$4.8\times10^{-48}\lesssim\frac{\Gamma(X\to\nu\gamma)}{M_X}\lesssim4.6\times10^{-46}$~\cite{Faisel:2014gda}.
\footnote{This bound is derived from $\sin^22\theta = (2-20)\times10^{-11}$~\cite{Baek:2014sda}.}
In our case, this form can be rewritten as
\begin{align}
4.8\times10^{-48}\lesssim
\frac{\alpha_{\rm em} M_X^2}{4(4\pi)^4}
\left|
\sum_{a}^{1,2}\sum_{j,k}^{1-3}\frac{m_{\ell_j} (f^\dag_a)_{jk}(g^\dag_a)_{j1} }{m^2_{h^\pm_a}}
\left(1-\frac{3-4\epsilon_{a,j}+2\ln\epsilon_{a,j} }{(1-\ln\epsilon_{a,j})^3} \right)
\right|^2
\lesssim4.6\times10^{-46},
\label{eq:x-ray}
\end{align}
where $\alpha_{\rm em}\approx1/137$ is the fine structure constant.


{\it Relic density}:
The observed relic density can be thermally generated  through the process of the GB final state due to the global $U(1)$ symmetry. 
The relevant Lagrangian is given by
\begin{align}
{\cal L}_{DM} = \frac{x}{6v'}(\partial_\mu G)(\bar N\gamma^\mu \gamma^5 N)+\frac1{v'} \sum_{b=1,2} V_{1b} (h_b \partial_\mu G \partial^\mu G).
\end{align}
Notice here that the term of $y_N$ cannot contribute to the annihilation process of  2$G$ final state, since this term does not
connect to the other terms with GB interactions due to the absence of the Lorentz index of $\mu$.
 Then the relativistic cross section of $X$  is given by
\begin{align}
\sigma v_{\rm rel}(s)\approx \frac{|V_{11}|^4 s (s-4M_X^2)}{256\pi v'^2(m_{h_1}^2 - s)^2},
\end{align}
where we neglect the contribution of the SM Higgs $h_2$.
Then the thermal averaged cross section is given by~\cite{Gondolo:1990dk}
\begin{align}
\langle\sigma v_{\rm rel}\rangle\approx
\frac{\int_{4 M^2_X}^\infty ds\sqrt{s-M^2_X}\sigma v_{\rm rel}(s) K_1(\sqrt{s}x/M_X)x}{16M_X^5 [K_2(x)]^2},
\end{align}
where $x\equiv M_X/T$ at the temperature $T$, and each of $K_{1,2}$ is the modified Bessel function of the second kind of order 1 and 2.
Finally the formula of the relic density ($\Omega_{X} h^2$) is approximately given by
\begin{align}
\Omega_{X} h^2\approx\frac{1.07\times 10^9 }{\sqrt{g_*(x_f)}M_{\rm Pl}\int_{x_f}^\infty dx\frac{\langle\sigma v_{\rm rel}\rangle}{x^2}[{\rm GeV}]},
\label{eq:relic}
\end{align}
where the lower index of $f$ represents the time of freeze-out, and each of $M_{\rm Pl}\approx1.22\times10^{19}$ GeV and $g_*(x_f)$
is the Planck mass and the number of the relativistic degrees of freedom. 
 To obtain the observed relic density $\Omega_{X} h^2\approx0.12$~\cite{Ade:2015xua}, we find a solution at $v'=1$ GeV and $m_{h_1}\approx0.051$ GeV,
where we use fixed reasonable values $|V_{11}|=1$, $x_f=20$, and $g_*(x_f)=100$ for brevity. We will adopt this solution in our numerical analysis.

\section{Numerical Analysis}
\begin{figure}[tbc]
\begin{center}
\includegraphics[scale=1.0]{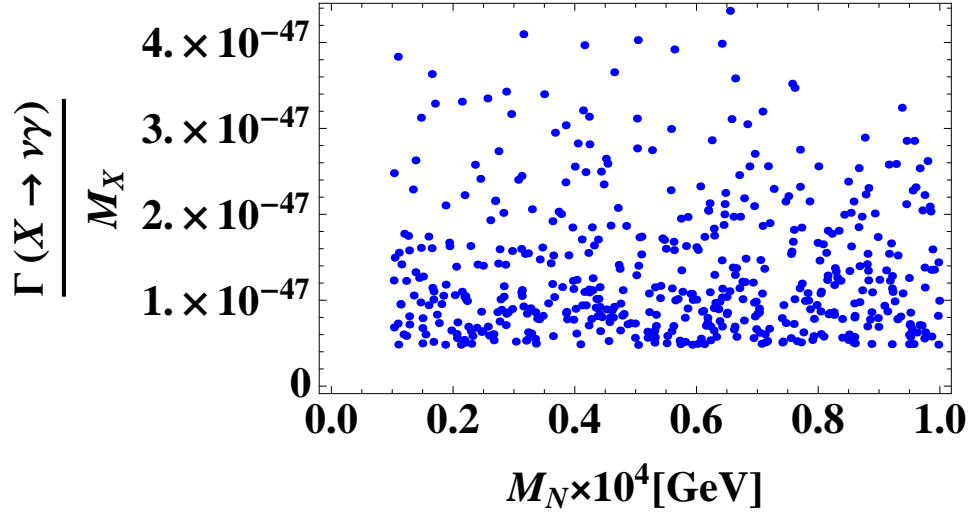}
\caption{Allowed scanning points in terms of $M_N$ and $\Gamma(X\to \nu\gamma)/M_X$, where these points satisfy all the constraints of LFV processes and lepton universalities as discussed in Section \ref{lfv-lu}.
}   \label{scan}
\end{center}
\end{figure}
In this section, we search the allowed region for Yukawa couplings $f$ and $g$.
First of all, we fix the following relevant parameters for simplicity
\begin{align}
& v'\approx1\ {\rm GeV},\ m_{h_1}\approx0.0506\  {\rm GeV},\ m_{h_2}\approx125 \ {\rm GeV},\ m_{h_1^\pm}\approx152 \ {\rm GeV},
\ m_{h_2^\pm} \approx193\  {\rm GeV}, \nn\\
&m_{\chi_1}\approx 124\  {\rm GeV},\ m_{\chi_2}\approx 121\  {\rm GeV},\  \nn\\
& M_X\approx 7.1\times10^{-6} \ {\rm GeV}, \ V\approx \left(\begin{array}{cc}
-1.00 & 1.54\times10^{-3}  \\ 1.54\times10^{-3}  & 1.00  \\\end{array}\right),\
O\approx \left(\begin{array}{cc}
-1.00 & 3.31\times10^{-3}  \\ 3.31\times10^{-3} & 1.00  \\\end{array}\right),\nn\\
& \lambda_\Phi\approx0.130,\  \lambda_{\varphi\Phi}\approx0.0984,\ \lambda_\varphi\approx0.254, \  \lambda_0\approx9.281,\
 \lambda_{\chi_1}\approx0.962,\   \lambda_{\chi_2}\approx0.614,\nn\\
& \lambda_{\chi_1\Phi}\approx0.864,\ \lambda_{\chi_2\Phi}\approx0.745,\ \lambda_{\chi_1\varphi}\approx0.0984,\ \lambda_{\chi_2\varphi}\approx0.915,\
\end{align}
where the condition of vacuum stability in Eq.~(\ref{eq:vs}) is satisfied. 
Now we can check whether the above input parameters satisfy the constraint of the invisible decay of SM-like Higgs in Eq.(\ref{eq:Binv}).
Our invisible decay width is computed as $\Gamma_{\rm inv}\approx0.0464$ that corresponds to the left-hand side of  Eq.(\ref{eq:Binv}).
On the other hand, the value of the right-hand side of  Eq.(\ref{eq:Binv}) is equal to $0.2$. Hence these input parameters satisfy this constraint. 

%
Then the order estimation of Yukawa couplings in Eq.(\ref{eq:yukawa-order}) reduces to $fg^T={\cal O}(10^{-4}-1)$, which can reproduce the sizable scale of neutrinos (0.01 eV).
Within this estimation, we randomly select fifteen values of Yukawa couplings as
\begin{align}
&  M_N\equiv M_{N_{2}}(=M_{N_{3}})\in [10^{-5},10^{-4}]\ [{\rm GeV}],\\
& f_{ij}\in[5\times 10^{-3},0.1], \\
& g_{ij}\in[5\times 10^{-3},0.1]\ {\rm for}\ (i,j)\neq \left((1,1),\ (1,2),\ (1,3)\right),\\
& g'\equiv g_{ij}\in[10^{-9},10^{-7}]\ {\rm for}\ (i,j)=\left((1,1),\ (1,2),\ (1,3)\right),
\end{align}
and we take 1000 scanning sample points over these above ranges. 
where we expect that a tiny value of $g'$ plays a role in explaining the X-ray line in order to satisfy Eq.(\ref{eq:x-ray}).  
Figure~\ref{scan} shows allowed scanning points(506) in terms of $M_N$ and $\Gamma(X\to \nu\gamma)/M_X$, where these points satisfy all the constraints of LFV processes and lepton universalities as discussed in Section \ref{lfv-lu}.

\section{ Conclusions and discussions}
We have constructed a two-loop induced neutrino model with a global $U(1)$ symmetry, and we have shown
allowed scanning points(506) in terms of $M_N$ and $\Gamma(X\to \nu\gamma)/M_X$ in Figure~\ref{scan}, where these points satisfy all the constraints of the scale of neutrino masses, LFV processes, lepton universalities, relic density of DM (at $v'=1$ GeV and $m_{h_1}\approx0.051$ GeV), and X-ray line  at 7.1 keV of the DM mass reported by XMN-Newton X-ray observatory using data of various galaxy clusters and Andromeda galaxy. 
Here DM does not link to the neutrino masses, since the DM mass is very tiny as well as $g'$ coupling.


As another phenomenological point of view, GB arises a discrepancy of the effective number of neutrino species in the early Universe, which is denoted by $\Delta N_{\rm eff}$. The recent data reported by Planck~\cite{Ade:2015xua} shows $\Delta N_{\rm eff}\lesssim0.04^{+0.33}_{-0.33}$ at the 95 \% confidential level. In our case, $\Delta N_{\rm eff}$ is about $0.045$, where we suppose that GB typically decouples from the plasma at temperatures just above the QCD phase transition(150-200 MeV) and our effective number of relativistic degrees of freedom is about 70.
Therefore, our model can evade this constraint. It is also worth mentioning that cosmological issues such as an effect on cosmic microwave background via cosmic string~\cite{Battye:2010xz} and stellar energy loss via the photoproduction process $\gamma+e\to e+$GB~\cite{Chang:1988aa} may put some constraints only when it couples to an axial vector current. However GB in our model has a vector current only, therefore we can evade these constraints, too.

\section*{ Appendix}
\begin{figure}[tbc]
\begin{center}
\includegraphics[scale=0.5]{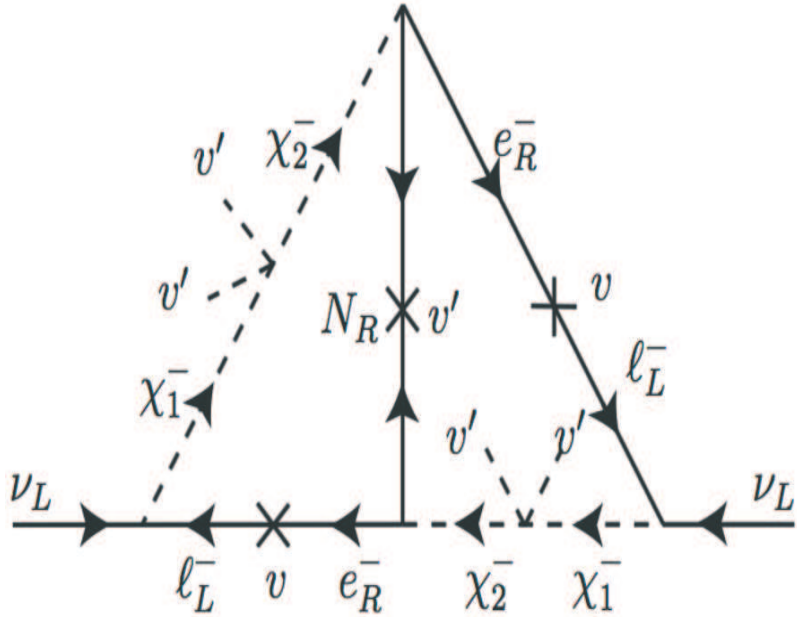}
\caption{ Subdominant contribution to the neutrino masses with Zee-Babu type.
}   \label{fig:neut2}
\end{center}
\end{figure}
The formula of the Zee-Babu type neutrino mass can be given by
\begin{align}
(m_\nu)_{nm}
=&-\sum_{a,b}^{1,2} \sum_{j,k,l}^{1-3}  \frac{f_{nj}^a m_{\ell_j}g_{kj}^b M_{N_k} g_{kl}^a m_{\ell_l}f_{lm}^b }{(4\pi)^4 M^2}\nn\\
\times&\int dxdydz\delta(x+y+z-1)\int dx'dy'dz'
\frac{\delta(x'+y'+z'-1)}{(y^2-y)y'X_{h_b^+}-(y X_{N_k}+z X_{h_a^+})z'},
\end{align}
where $X_f\equiv (m_f/M)^2$, and $M\equiv {\rm Max} (m_{h_a^+},m_{h_b^+},M_{N_k})\approx{\rm Max} (m_{h_a^+},m_{h_b^+})$, as shown in Figure~\ref{fig:neut2}.
In our scanning range of the numerical analysis, the typical scale of the contribution to the neutrino masses is  around ${\rm O}(10^{-9})$ eV that is negligible.

\vspace{0.3cm}
{\it Acknowledgments}:

Author thanks to Prof. Seungwon Baek, Dr. Kenji Nishiwaki, Dr. Yuta Orikasa, Dr. Takashi Toma, and Dr. Kei Yagyu for fruitful discussions.

\end{document}